\documentclass{article}
\usepackage{epsfig}

\date{\today}

\begin{document}
\begin{center}
{\Large\bf NEW HAIRY BLACK HOLES WITH NEGATIVE COSMOLOGICAL CONSTANT}
\vspace{1.2cm}\\
J.J.\ van der Bij and Eugen Radu\footnote{\textbf{corresponding author}:
\\Albert-Ludwigs-Universit\"at Freiburg, Fakult\"at f\"ur Physik, 
\\Hermann-Herder-Stra\ss e 3, D-79104 Freiburg, Germany
\\ email: radu@newton.physik.uni-freiburg.de
\\ Telephone: +49-761/203-7630 
\\Fax: +49-761/203-5967}\vspace{0.4cm}\\
\it Albert-Ludwigs-Universit\"at \\
\it Fakult\"at f\"ur Physik\\
\it Freiburg Germany\vspace{1.2cm}\\
\end{center}

\begin{abstract}
Black hole solutions  with nonspherical event horizon topology
are shown to exist 
in an Einstein-Yang-Mills theory 
with negative cosmological constant.
The main characteristics of the solutions are presented and  
differences  with respect to the  spherically symmetric case are studied.
The stability of these configurations is also addressed.
\end{abstract}
\\
\\
Although anti-de Sitter (AdS) spacetime does not seem to correspond to the world
in which we live \cite{Carroll:2001fy}, its importance has been noticed in many occasions.
As shown by Hawking and Page, the presence of a negative comological constant $\Lambda$
makes it possible for a black hole to reach stable thermal equilibrium 
with a heat bath \cite{Hawking:1983dh}. 

The black holes discussed by Hawking and Page have a spherically symmetric 
event horizon. 
The topological structure of the event horizon of a black hole is an intriguing 
subject in black hole physics.
When asymptotic flatness and the energy conditions are given up, there
are no fundamental reasons to forbid the existence of black holes 
with nontrivial topologies.
In particular, for a negative cosmological constant, 
black holes for which the topology of the horizon 
is an arbitrary genus Riemann surface have been considered
by many authors (see $e.g.$ \cite{Mann:1997gj}-\cite{Cai:1998ii}). 
The thermodynamics of these solutions has been discussed 
in \cite{Brill:1997mf}-\cite{Peca:2000dv} while
higher dimensional generalizations were obtained in \cite{Mann:1997iz}-\cite{Aros:2001ij}.
The theorems about spherical horizon topology do not apply here because
the negative cosmological constant can be interpreted 
as a negative vacuum energy density.
They generalize the known solutions replacing the round 
two-sphere by a two-dimensional space $\Sigma$ 
of negative or vanishing curvature.

These black holes embedded in 'locally AdS' background spacetimes (background locally
isometric to spacetimes of constant negative curvature) have been seminal 
to recent developments in black hole physics.

All these investigations are mainly based on Einstein(-Maxwell) theory 
(note however the inclusion of a dilaton field in \cite{Cai:1998ii}).
It may be of interest to generalize these solutions for a nonabelian
matter content.
Here we restrict attention to $SU(2)$ Einstein-Yang-Mills (EYM)
theory and investigate black hole 
solutions with locally flat or hyperbolic horizons, which asymptotically
approach a locally AdS spacetime.

The case of a spherically symmetric horizon has been discussed 
in \cite{Winstanley:1999sn,Bjoraker:2000qd} with
some surprising results.
For example, there are finite energy solutions for 
several intervals of the shooting parameter 
(the value of gauge function at event horizon), rather than discrete values. 
Solutions exist for all values of $\Lambda<0$.
There are also stable solutions in which the gauge field has no zeros, and the asymptotic
value of the gauge potential is not fixed.

This behavior is drastically different
from those observed for EYM black holes in asymptotically flat \cite{Bizon:1990sr}
or de Sitter spacetime \cite{Torii:1995wv}.
Their regular conterparts are discussed in \cite{Bjoraker:2000qd, Bjoraker:2000yd}
and present also interesting properties.

In this letter we show numerically that,
in the EYM system, the topology of the event horizon 
is not restricted to the spherically symmetric case.
The family of black hole solutions considered in 
\cite{Winstanley:1999sn,Bjoraker:2000qd} 
is extended to geometries with locally flat and hyperbolic horizons.

We consider spacetimes whose metric can be written locally in the form
\begin{equation}\label{metric}
ds^{2}=\frac{dr^{2}}{H(r)}+r^2d \Omega_k^2-
\frac{H(r)}{p^2(r)} dt^{2}
\end{equation}
where
\begin{equation} 
H(r)=k-\frac{2m(r)}{r}-\frac{\Lambda r^2}{3}.
\end{equation}
Here $d \Omega_k^2=d\theta^{2}+f^{2}(\theta) d\varphi^{2}$
is the metric on a two-dimensional surface $\Sigma$ of constant curvature $2k$.
$r$ is the radial coordinate for which 
$r\to \infty$ defines the asymptotic region.
The discrete parameter $k$ takes the values $1, 0$ and $-1$ 
and implies the form of the function $f(\theta)$
\begin{equation}
f(\theta)=\left \{
\begin{array}{ll}
\sin\theta, & {\rm for}\ \ k=1 \\
\theta , & {\rm for}\ \ k=0 \\
\sinh \theta, & {\rm for}\ \ k=-1.
\end{array} \right.
\end{equation}
In case of vanishing $\Lambda$, these metrics describe black holes 
only for spherically symmetric solutions.
The topology of spacetime is $R^2\times H^2_g$, 
where $H^2_g$ is the topology of the surface $\Sigma$.
When $k=+1$, the universe takes on the familiar spherically symmetric form,
and the $(\theta, \varphi)$ sector has constant positive curvature. 
The cosmological constant may assume either sign. 
When $k=0$, the $\Sigma$ is a flat surface, with $\Lambda$ less than zero.
When $k=-1$, $\Lambda<0$ and 
the $(\theta, \varphi)$ sector is a space with constant negative curvature, 
also known as a hyperbolic plane. 
A discussion of these cases is available in \cite{Brill:1997mf}.

When $\Sigma$ is closed, we denote its area by $V$.
 
The EYM coupled system is described by the action
\begin{equation} \label{action}
S=\int d^{4}x\sqrt{-g}[\frac{1}{16\pi G}(\mathcal{R} - 2 \Lambda)
-\frac{1}{4}F_{\mu \nu }^{a} F^{a\mu \nu }], 
\end{equation}
 
The most general expression for the appropriate $SU(2)$ connection 
is obtained by using the standard rule 
for calculating the gauge potentials for any spacetime group 
\cite{Forgacs:1980zs}.
Taking into account the symmetries of the line element (\ref{metric}) we find 
\begin{equation} \label{A}
A=\frac{1}{2e} \{ u(r,t) \tau_3 dt+ \nu(r,t) \tau_3 dr+
\left( \omega(r,t) \tau_1 +\tilde{\omega}(r,t) \tau_2\right) d \theta
+\left(\frac{d \ln f}{d \theta} \tau_3
+ \omega(r,t) \tau_2-\tilde{\omega}(r,t)\tau_1  \right) f d \varphi \}, 
\end{equation}
 
where $\tau_a$ are the Pauli spin matrices.
To simplify the general picture we set $u=0$ $i.e.$ no dyons.
However, nontrivial solutions exist also for a nonvanishing electric part 
(see \cite{Bjoraker:2000qd} for a discussion of black hole dyon solutions in the $k=1$ case). 
The existence of dyon solutions without a Higgs field
is a new feature for AdS spacetime;
if $\Lambda \ge 0$ the electric part of the gauge fields
is forbidden \cite{Bjoraker:2000qd, Galtsov:1989ip}. 

For static configurations, it is convenient to take the $\nu=0$ gauge
and eliminate $\tilde{\omega}$ by using a residual gauge freedom. 
The remaining function $\omega$ depends only on the coordinate $r$.

As a result, we obtain a simplified YM curvature $F=dA-ieA \wedge A$
\begin{equation} 
F=\frac{1}{2e}\left (
\omega' \tau_1 dr\wedge d\theta +
f \omega' \tau_2 dr\wedge d\varphi +
(w^2-k)f \tau_3 d\theta \wedge d\varphi \right )
\end{equation}
where a prime denotes a derivative with respect to $r$.
For this purely magnetic YM ansatz, we obtain the equations
\begin{eqnarray} 
m'&=&\omega'^2 H+\frac{(\omega^2-k)^2}{2 r^2},\label{e1}
\\
\omega''&=&\frac{\omega'}{H} \left ( \frac{2\Lambda r}{3}
-\frac{2m}{r^2}+\frac{(\omega^2-k)^2}{r^3} \right)
+\frac{\omega (\omega^2-k)}{r^2 H}, \label{e2}
,
\end{eqnarray}
 
where we have used dimensionless variables rescaled as
$r \to (\sqrt{4\pi G}/e) r,~\Lambda \to (e^2/4 \pi G) \Lambda$
and $m \to (eG/\sqrt{4\pi G}) m$.
\newline
The equation for $p$ decouples from the rest. 
Given a solution to (\ref{e2}), $p$ is obtained from
\begin{eqnarray} \label{p}
p=\exp \left (2 \int_r^{\infty} dr \frac{\omega'^2}{r} \right ).
\end{eqnarray} 
For black hole solutions having a regular event horizon at $r=r_h>0$,
\begin{eqnarray}
m(r_h)=\frac{r_h}{2}\left(k-\frac{\Lambda r_h^2}{3}\right).
\end{eqnarray}
We find the following expansion near the event horizon
\begin{eqnarray} \label{expansion}
m(r)&=&m(r_h)+m'(r_h)(r-r_h),\label{h1}
\\
\omega(r)&=&\omega_h+\omega'(r_h)(r-r_h),\label{h3}
\end{eqnarray}
 
where
\begin{eqnarray} 
m'(r_h)&=&\frac{(\omega_h^2-k)^2}{2 r_h^2},
\\
\omega'(r_h)&=&\frac{r_h\omega_h(\omega_h^2-k)}
{(k-\Lambda r_h^2)r_h^2-(\omega_h^2-k)^2}.
\end{eqnarray}
 
The condition for a regular event horizon is
\begin{eqnarray}
\left.
\frac {d}{dr} \left( 
k-\frac {2m}{r} -\frac {\Lambda r^{2}}{3} \right) 
\right| _{r_{h}} >0.
\end{eqnarray}
For given ($r_h,~\Lambda$), this places a bound on $\omega_h$
\begin{eqnarray}\label{mhbound}
2m'(r_h ) =\frac {\left( \omega_h^{2}-k \right) ^{2}}{r_h ^{2}}
<k-\Lambda r_h ^{2},
\end{eqnarray}
and implies the positiveness of the quantity $\omega'(r_h)$.
In the $k=-1$ case, (\ref{mhbound}) implies the existence of 
a minimal value of $|\Lambda|$, $i.e.$ for a given $r_h$
\begin{eqnarray}
|\Lambda|>\frac{1}{r_h^2}(1+\frac{1}{r_h^2}).
\end{eqnarray}

At $r \to \infty$, the spacetime is locally isometric to AdS spacetime.
In this limit, the field  variables 
have the following asymptotic forms
\begin{eqnarray}
m(r) & = & M+\frac {M_{1}}{r} +O\left( \frac {1}{r^{2}} \right) \label{b1},
\\
\omega (r) & = & \omega _{\infty } +\frac {C_1}{r} 
+O\left( \frac {1}{r^{2}} \right)\label{b2},
\end{eqnarray}
where  $\omega _{\infty }$, $M$ and $C_1$ are parameters and
$
M_{1} = \frac {\Lambda C_1^{2}}{3}-\frac {1}{2} ( 
\omega _{\infty }^{2}-k) ^{2}.
$
For $k=1$, $m(r)$ is the mass contained inside the radius $r$, 
while $M$ is the Arnowitt-Deser-Misner (ADM) mass.  
In the general case, $M$ is related to mass of the hole $\mathbf{M}_{ADM}$
(appropriately generalized for a negative 
cosmological constant \cite{Abbott:1982ff}) by
$\mathbf{M}_{ADM}=MV/4\pi$.
However, for $k=-1$ this identification is contentious.
At issue is the zero of mass-energy, since even in pure Einstein gravity 
(with negative $\Lambda$) one has a black hole solution with $M<0$.
Surprisingly, this negative mass black hole 
can also be formed by gravitational collapse \cite{Mann:1997jb}.
Vanzo \cite{Vanzo:1997gw} gives explicit mass 
formul\ae\ both for the solutions in an EM theory
and for more general stationary black holes with the same asymptotic 
behaviour.
In Ref. \cite{Vanzo:1997gw}, the shifting of the zero of energy 
is advocated so that the total energy is
proportional to
$M-M_0$, assuming there is suitable shift $M_0$ that works not only for
the special solutions considered by Vanzo, but for all reasonable
solutions with the same asymptotic behaviour.

The black hole solutions discussed in this paper contain 
a YM magnetic charge
\begin{eqnarray} 
\mathbf{Q_M}=\frac{1}{4 \pi}\int dS_k \sqrt{-g} \tilde{F}^{kt}=
Q_M \frac{V}{4 \pi}\frac{\tau_3}{2},
\end{eqnarray}
 
where $Q_M=(k-\omega_{\infty }^2)$. 
We remark that all $k=-1$ black holes posses a nonzero magnetic charge.
$\mathbf{M}_{ADM}$ and $\mathbf{Q_M}$ turn out to be finite if
$\Sigma$ is closed.

We numerically solve Eqs. (\ref{e1}) and(\ref{e2})
with boundary conditions (\ref{b1}), (\ref{b2})
using a standard shooting method. 
As mentioned above, the case $k=1$ is discussed in 
Ref. \cite{Winstanley:1999sn,Bjoraker:2000qd}.
We shall therefore focus on the cases $k=0$ and $k=-1$.

Using the initial conditions on the event horizon
(\ref{h1}), (\ref{h3}), the equations were integrated 
for a range of values of $(r_h,~\Lambda)$ 
and varying $\omega _{h}$.
Since the equations (\ref{e1}), (\ref{e2}) are invariant under the
transformation $\omega \rightarrow - \omega $, only values of
$\omega _{h}>0$ are considered.
In the $k=-1$ case, for $\omega_h=0$, $\omega'(r_h)=0$ and we have 
the corresponding EM solutions
discussed in \cite{Brill:1997mf}.
For sufficiently small $\omega_h$, all field variables remain 
close to their values for the abelian configuration
 with the same $r_h$.
Significant differences occur for large enough values of $\omega_h$
and the effect of the nonabelian field on the geometry 
becomes more and more pronounced.
A similar behavior is noticed for $k=0$; here $\omega_h=0$ corresponds to
a vacuum solution.

The solutions we obtain have many common properties with their
spherically symmetric $\Lambda<0$ counterparts.
Also, most of the analytical results found in 
\cite{Winstanley:1999sn,Bjoraker:2000qd} for $k=1$ 
can easily be generalized for the cases $k=0,-1$.
As expected, the black holes properties are rather different from the 
corresponding ones for asymptotically flat and de Sitter spacetimes.

Solutions are classified by the parameters $M$ and $Q_M$.
The behavior of metric functions $m$ and $p$ is similar to the spherically symmetric 
black hole solutions.
For every considered value of $(\Lambda,~r_h)$, we find regular black hole solutions
for only one interval $0<\omega_h< \omega_h^c$.
For $\omega_h>\omega_h^c$ solutions blow up or the function $H(r)$ becomes negative.
The value of $\omega_h^c$ increases as $|\Lambda |$ increases.
There are also solutions for which $\omega_{\infty}>1$
although $\omega_h<1$.

In contrast to the spherically symmetric case,
we find only nodeless solutions. 
This can be analytically proven by 
integrating the equation for $\omega$, 
$(H\omega'/p)'=\omega(\omega^2-k)/p r^2$ between $r_h$ and $r$; 
thus we obtain $\omega'>0$ for every $r>r_h$.
For $k=1$, 
both nodeless solutions and solutions where $\omega$ crosses the axis can exist.

For $k=0$ we have always $m>0$ and the black holes 
therefore only occur with positive values
of mass. Typical solutions in this case are presented in Figure 1.
In Figure 2, $M$ is plotted as a function of $Q_M$ for $k=0$ black holes
and several values of $\Lambda$.
 
Not unexpectedly, when $\Sigma$ has a negative curvature,
we have found for suitable values of ($\Lambda,~r_h,~\omega_h$),
a class of solutions with $M<0$ or even $M=0$, 
apart from configurations with positive $M$ (see Figure 3).
Typical monopole spectra for $k=-1$ case are presented in Figure 4.

Black hole solutions in
asymptotically Minkowski and de Sitter space, which necessarily have at
least one node in $w(r)$,  are unstable
\cite{Straumann:1990as, Torii:1995wv}. 
In contrast, the spherically symmetric monopole black hole 
solutions in the asymptotically  
AdS spacetime with no node in $\omega(r)$ are stable 
\cite{Winstanley:1999sn},\cite{Bjoraker:2000qd},\cite{Sarbach:2001mc}.

We shall show that
there are also black hole solutions with nonspherical event horizon
stable against linear fluctuations
(this is likely since we have found only nodeless solutions).

In examining time-dependent fluctuations around black hole solutions it is
convenient to work in the $u(r,t)=0$ gauge.  
All field variables are written as the sum of the static 
equilibrium solution whose stability we are investigating
and a time dependent perturbation.
By following the standard methods well-known from the  
spherically symmetric case,
we derive linearized equations for $\delta w(r,t)$, $\delta \tilde
w(r,t)$, $\delta \nu(r,t)$, $\delta p(r,t)$ and $\delta m(r,t)$. 
The fluctuations decouple in two groups:   
$\delta w(r,t)$,   $\delta p(r,t)$ and $\delta m(r,t)$
form the gravitational sector, whereas $\delta \tilde w(r,t)$ and 
$\delta \nu(r,t)$ form matter (sphaleronic) sector.  The linearized 
equations imply that  $\delta p(r,t)$ and $\delta m(r,t)$ are
determined by $\delta w(r,t)$, and  $\delta \tilde w(r,t)$
 by $\delta \nu(r,t)$.
For a harmonic time dependence $e^{i \Omega t}$, the linearized system
of the matter sector implies a standard Schr\"odinger equation  
\begin{eqnarray}\label{sph-perturb}
\left \{-\frac{d^2}{d \rho^2}
+U_{\beta}(\rho) \right \}\beta&=&\Omega^2 \beta,
\end{eqnarray}
where $\beta(r, t)=r^2 p \delta \nu/\omega=e^{-i \Omega t}\beta(r)$ 
and a new tortoise radial coordinate is introduced $d \rho/dr=p/H$.
\newline
Here 
\begin{eqnarray}\label{pot1}
U_{\beta}&=&\frac{H}{r^2 p^2}(\omega^2+k)
+\frac{2}{\omega^2}(\frac{d \omega}{d \rho})^2,
\end{eqnarray}
with $U_{\beta}(r_h)=0$.
Since $\omega$ has no nodes, this potential is regular everywhere.
For $k=0$, $U_{\beta}>0$ and the potential is greater than the lower of 
its two asymptotic values.
Standard results from quantum mechanics 
\cite{Messiah, Winstanley:1999sn} imply that 
there are no negative eigenvalues for $\Omega^2$ and no unstable modes.
The same condition is satisfied by those $k=-1$ solutions with $\omega_{\infty}>1$. 
\newline
Similarly, the gravitational sector perturbation satisfies the equation
\begin{eqnarray}\label{grav-perturb}
\left \{-\frac{d^2}{d \rho^2}+U_{\omega}(\rho) \right \}\delta \omega
&=&\Omega^2 \delta \omega,
\end{eqnarray}
with the potential
\begin{eqnarray}\label{pot2}
U_{\omega}&=&\frac{H}{r^2 p^2}(3\omega^2-k)
+4\frac{d}{d \rho}\left [ \frac{p}{rH}(\frac{d \omega}{d \rho})\right]
\end{eqnarray}
or, by using the equations (\ref{e1})-(\ref{e2})
\begin{eqnarray}\label{pot2-1}
U_{\omega}&=&\frac{H}{r^2 p^2}\left [
3\omega^2-k+\frac{8 \omega \omega'}{r}(\omega^2-k)+4\Lambda r^2\omega'^2
-4k\omega'^2+\frac{4\omega'^2(\omega^2-k)^2}{r^2}
\right].
\end{eqnarray}
This potential is not positive definite, 
but is regular in the entire range $-\infty<\rho< \infty$.
Near the event horizon, $U_{\omega} \to 0$; at infinity $U_{\omega}$ 
takes always a positive value.
It follows that eq. (\ref{grav-perturb}) will have no bound states if the
potential $U_{\omega}$ is everywhere greater than the lower of its two asymptotic
values $i.e.~U_{\omega}>0$.
We have found numerically $U_{\omega}>0$ for all $k=0$ considered configurations.
This condition holds also for all $k=-1$ solutions stable in the sphaleronic sectors
(with $\omega_{\infty}>1$).
Thus, these solutions are linearly stable in both sectors of the theory.

This fact can also be proven more rigorously.
The analytical arguments presented in \cite{Winstanley:1999sn}
for the existence
of linearly stable black hole solutions 
can be directly generalized for every $k$. 
The only negative term in (\ref{pot2-1}) is $4\Lambda r^2\omega'^2$.
Following \cite{Winstanley:1999sn} we can show that 
this term can be made as small as we like by 
taking $|\Lambda|$ sufficiently large.
Thus we see that a nonspherical topology of the event horizon
has a tendency to stabilize the unperturbed solution.

The thermodynamic stability of these black holes will not be considered here and
we will restrict ourselves on deriving the Hawking temperature.
This can easily be done by using the standard Euclidean method.
For the line element (\ref{metric}), if we treat $t$ as complex, 
then its imaginary part is a coordinate for a nonsingular 
Euclidean submanifold iff it is periodic with period 
\begin{eqnarray}\label{period}
\beta=\frac{4 \pi r_h p(r_h)}{k-2m'(r_h)-\Lambda r_h^2}.
\end{eqnarray}
Then continuous Euclidean Green functions must have this period, so by standard
arguments the Hawking temperature is (with $k_B=\hbar=1$)
\begin{eqnarray}\label{Temp}
T_H=\frac{\left( k-2m'(r_h)-\Lambda r_h^2 \right )}{4 \pi r_h p(r_h) } 
\leq \frac{k-\Lambda r_h^2}{4 \pi r_h}.
\end{eqnarray}
This is a general result, for every matter content 
satisfying the weak energy condition.
Thus the Hawking temperature of such systems appears to be suppressed relative
to that of a vacuum black hole of equal  horizon area.
This relation generalizes for a negative cosmological constant the result
found in \cite{Visser:1992qh} for a vanishing $\Lambda$.
Note also that, for given $k$, $T_H$ is independent 
of the topology of two-space $\Sigma$.

In Figure 5 we plot the temperature $T_H$ $vs.$ the parameter $M$ for black holes
with fixed ($\Lambda,~r_h$) and varying the parameter $\omega_h$.
On these curves, the maximal values of $T_H$ correspond to $\omega_h=0$. 

The configurations we have found broaden the family of 
black holes with nonabelian matter content and manifest some new features.
Also, all known hairy EYM black hole solutions have particle-like, regular counterparts.
However, there are no regular counterparts of 
the solutions discussed in this work.
This fact may be of some relevance 
when generalizing the isolated horizon formalism 
for a negative cosmological constant \cite{Corichi:2001ct}.
We conjecture the existence of nontrivial black hole solutions with 
nonspherical event horizon for a more general nonabelian matter 
content $e.g.$ when including a Higgs field.
\\
\newline
{\bf Acknowledgement}
\newline
We gratefully thank Dr. B. Tausk for a careful reading of this manuscript.
\\
 This work was performed in the context of the
Graduiertenkolleg of the Deutsche Forschungsgemeinschaft (DFG):
Nichtlineare Differentialgleichungen: Modellierung,Theorie, Numerik, Visualisierung.


\newpage
{\bf Figure Captions}
\newline

Figure 1:
The gauge function $\omega$
and the mass $m$ are shown as a function of the radial coordinate
$r$ for three  different solutions; here $k=0$, $r_h=1$.
\newline

Figure 2:
Black hole monopole spectrum for $k=0$ black hole solutions with $r_h=1$ 
and different values of $\Lambda$.
\newline

Figure 3:
The gauge function $\omega$
and the mass $m$ are shown as a function of the radial coordinate
$r$ for three  different solutions; here $k=-1$, $r_h=5$.
\newline

Figure 4:
Black hole monopole spectrum for $k=-1$ black hole solutions with $r_h=5$,
and different values of $\Lambda$.
The existence of solutions with negative values of $M$ for small values of 
parameter $\omega_h$ is explicit.
\newline

Figure 5:
The variation of the Hawking temperature $T_H$ with $M$ for topological black holes
as a function of the parameter $\omega_h$. 
The radius of event horizon 
radius $r_h$ is fixed and several values of 
cosmological constant are considered.
On these curves, the maximal values of $T_H$ correspond 
to $\omega_h=0$.
\newpage

\setlength{\unitlength}{1cm}

\begin{picture}(16,16)
\centering
\put(-2,0){\epsfig{file=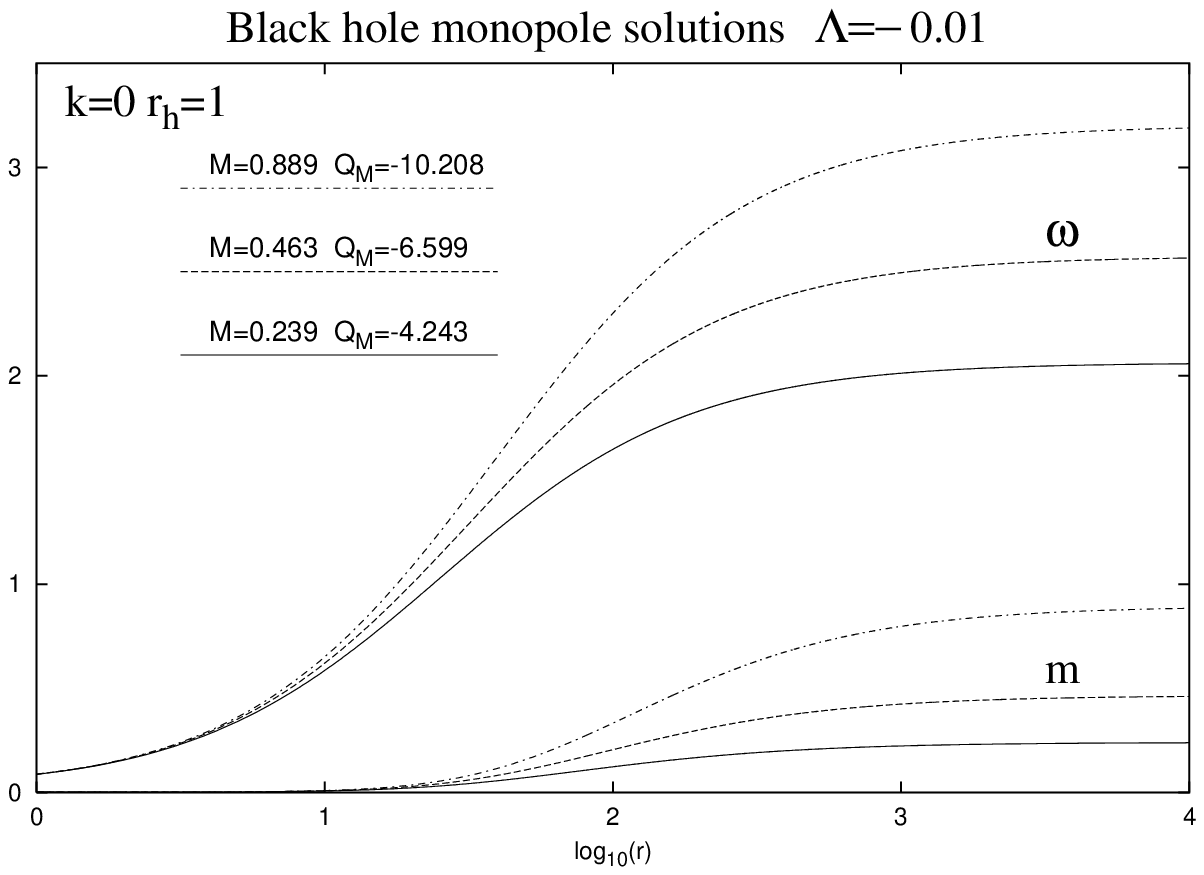,width=16cm}}
\end{picture}
\begin{center}
Figure 1
\end{center}

\newpage
\setlength{\unitlength}{1cm}

\begin{picture}(16,16)
\centering
\put(-2,0){\epsfig{file=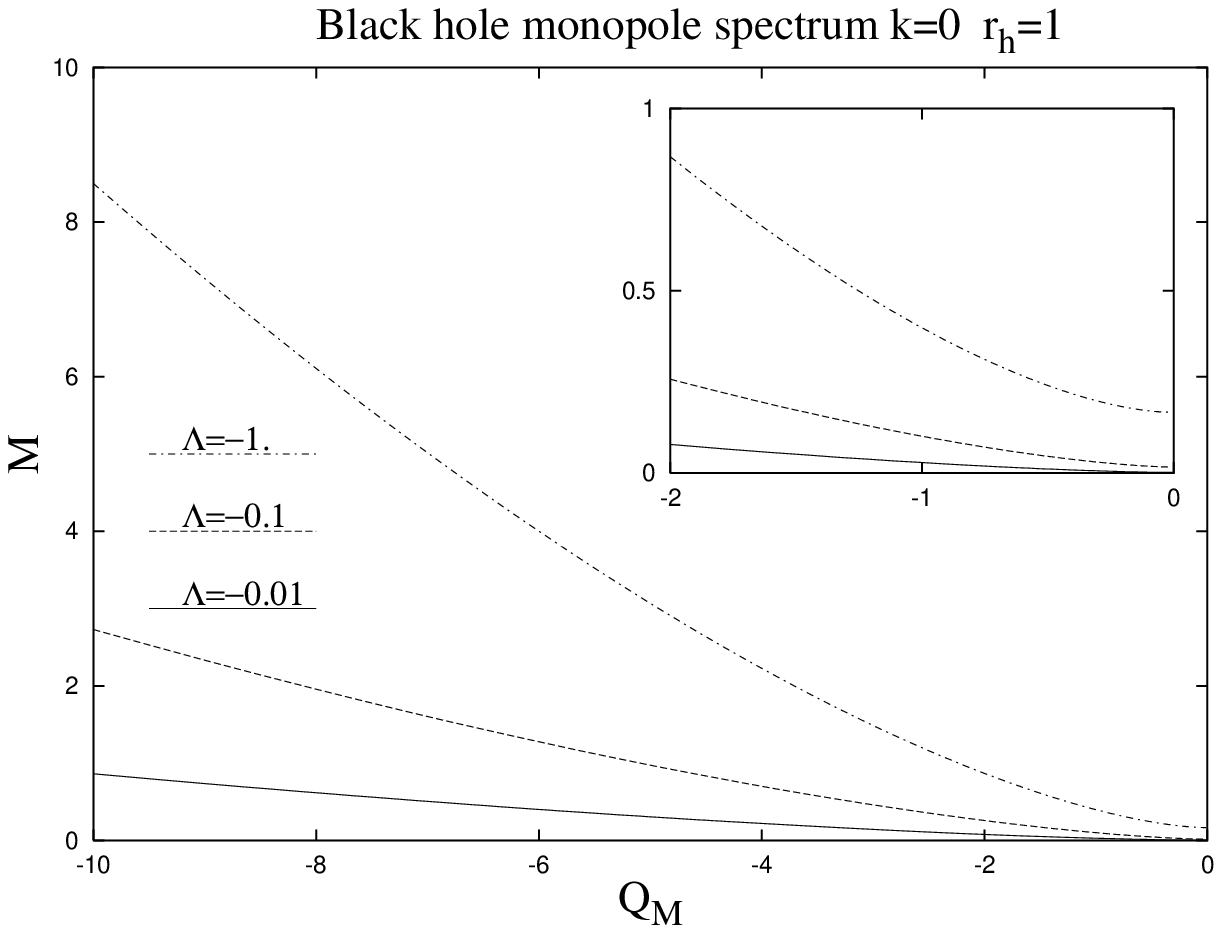,width=16cm}}
\end{picture}
\begin{center}
Figure 2.
\end{center}

\newpage
\setlength{\unitlength}{1cm}

\begin{picture}(16,16)
\centering
\put(-2,0){\epsfig{file=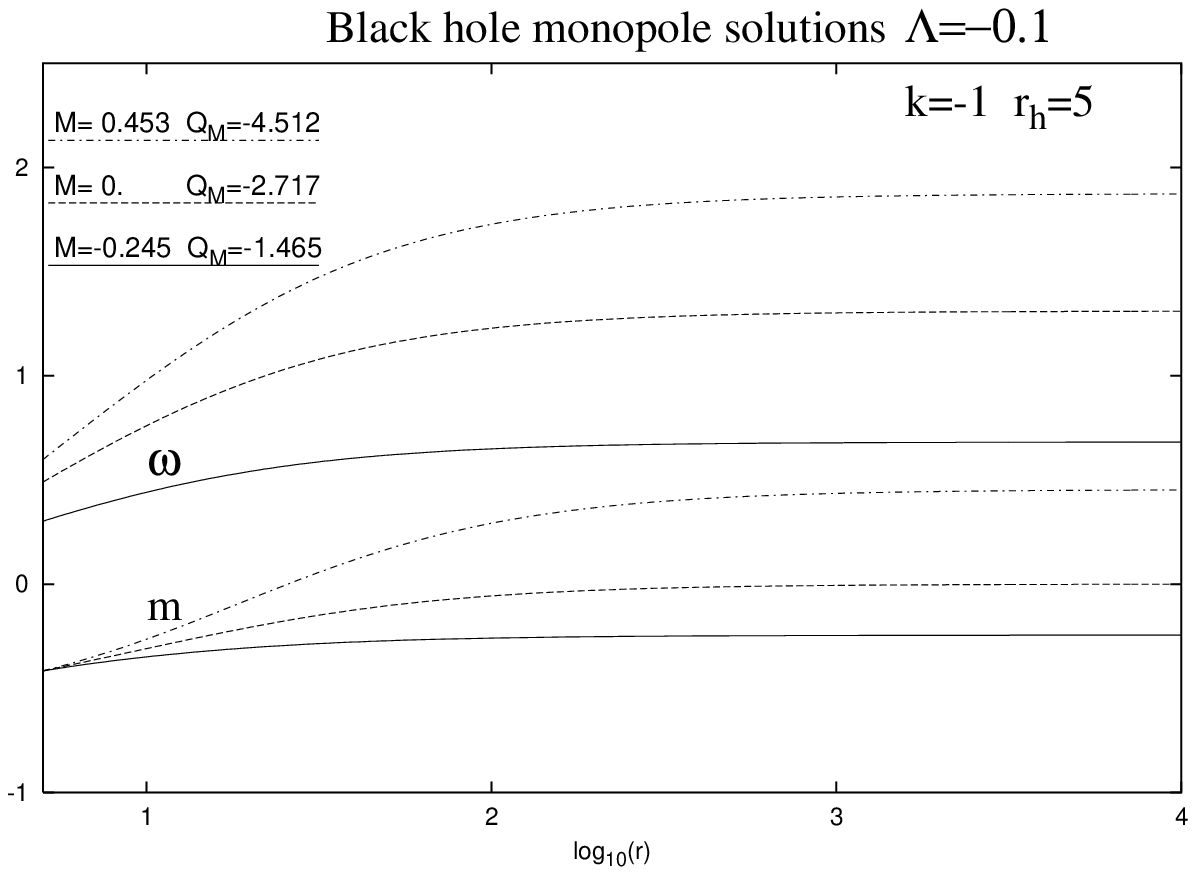,width=16cm}}
\end{picture}
\begin{center}
Figure 3.
\end{center}

\newpage
\setlength{\unitlength}{1cm}

\begin{picture}(16,16)
\centering
\put(-2,0){\epsfig{file=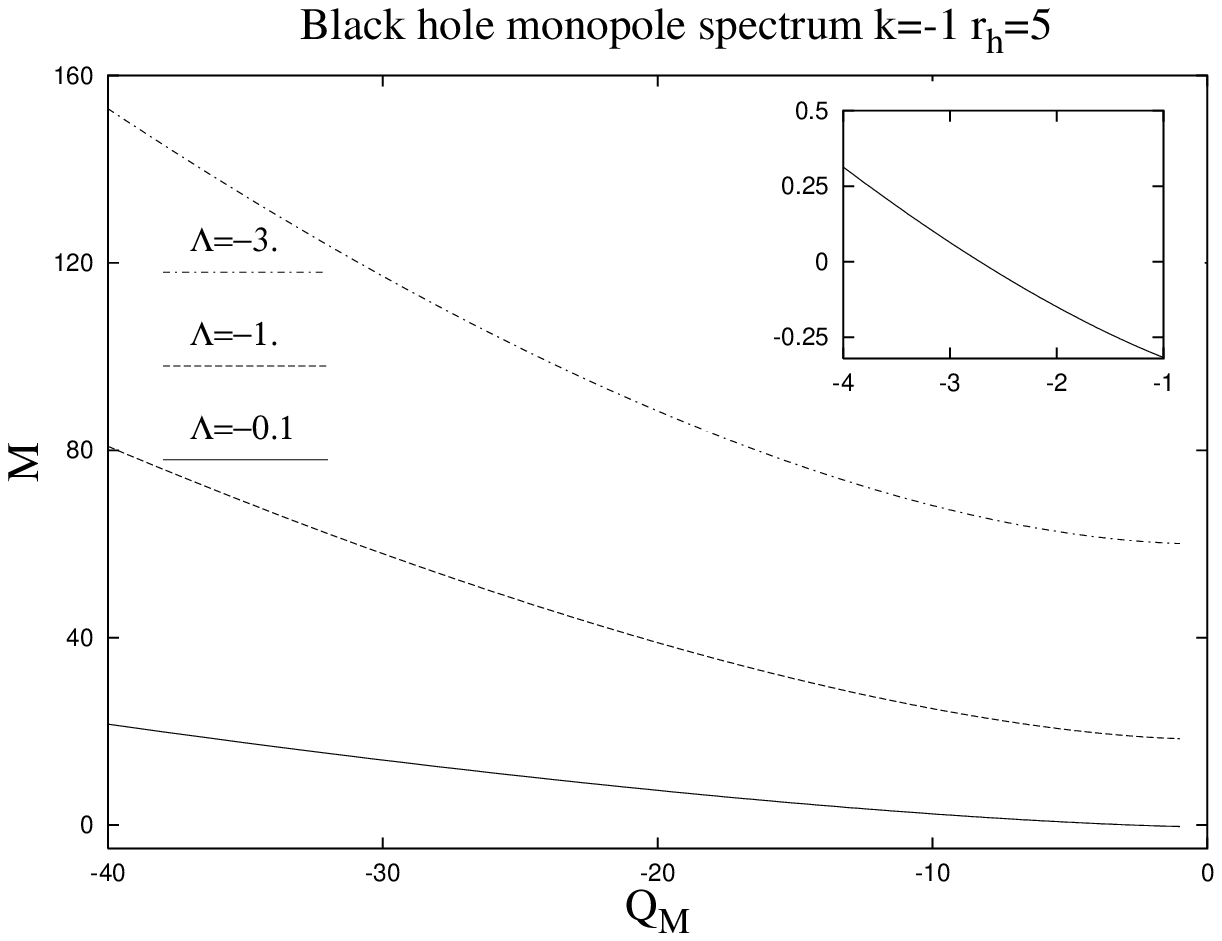,width=16cm}}
\end{picture}
\begin{center}
Figure 4.
\end{center}

\newpage
\setlength{\unitlength}{1cm}

\begin{picture}(16,16)
\centering
\put(-2,0){\epsfig{file=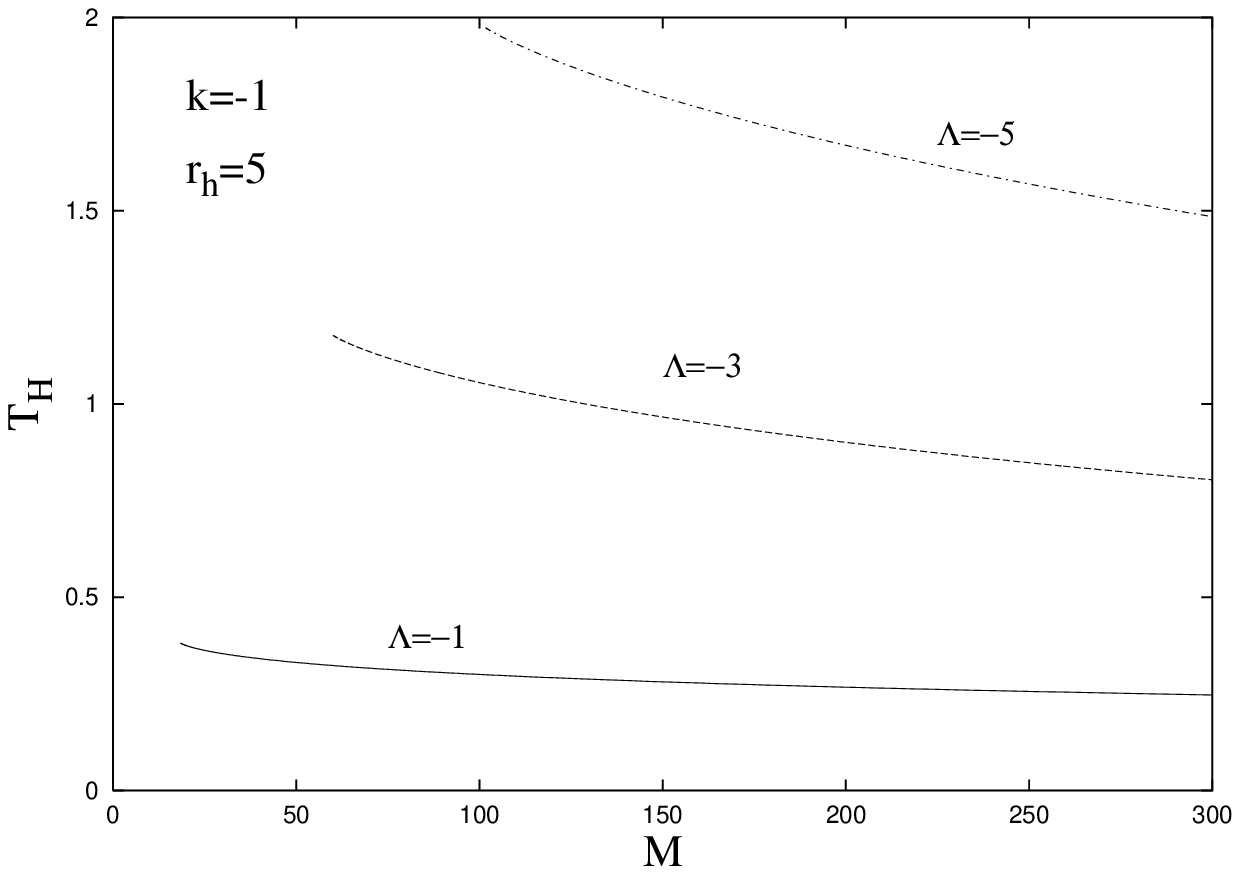,width=16cm}}
\end{picture}
\begin{center}
Figure 5a.
\end{center}

\newpage
\setlength{\unitlength}{1cm}

\begin{picture}(16,16)
\centering
\put(-2,0){\epsfig{file=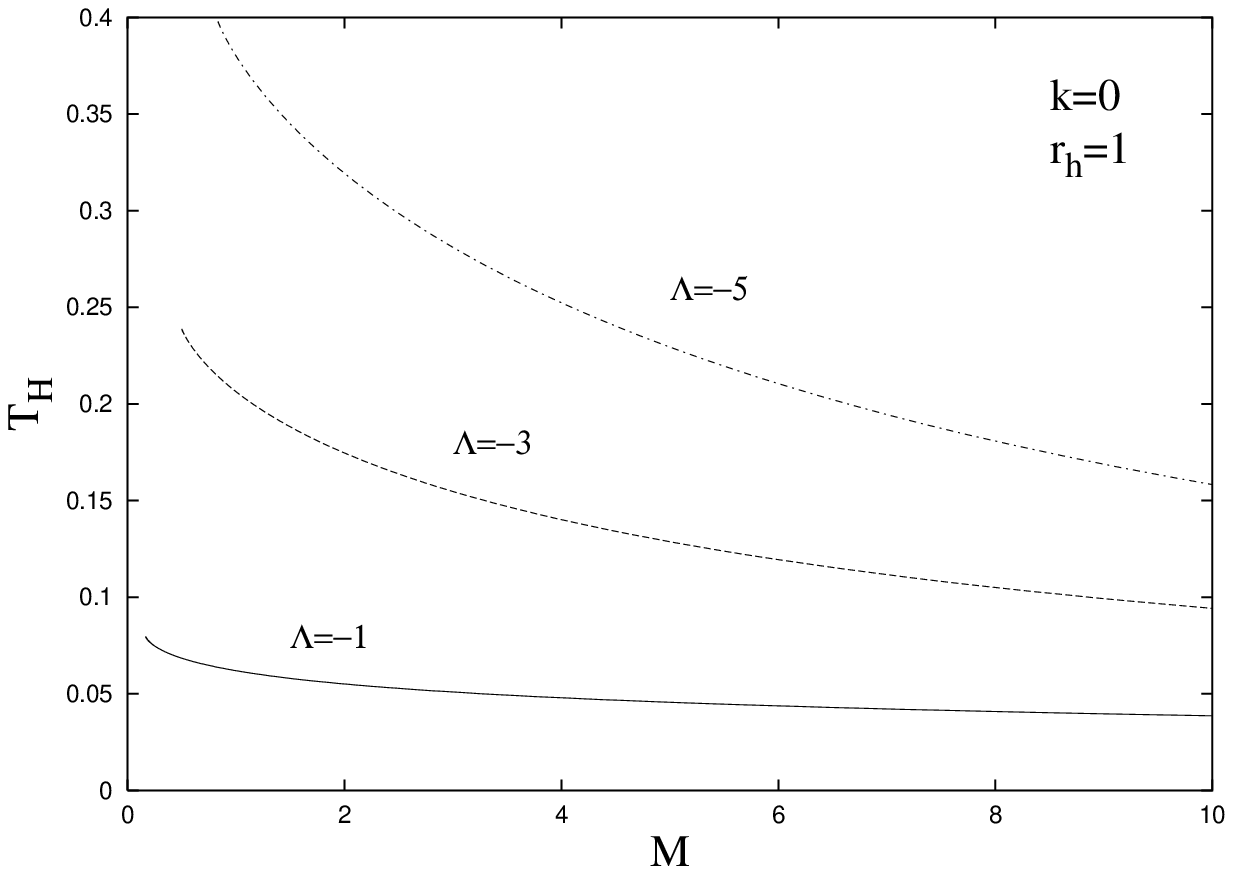,width=16cm}}
\end{picture}
\begin{center}
Figure 5b.
\end{center}
\end{document}